\documentclass[12pt]{article}
\usepackage{graphicx}
\usepackage{graphics}
\input epsf

\begin{document}

\title{\Large \bf ACORDE a Cosmic Ray Detector for ALICE}
\date{ }

\author{\large A.~Fern\'andez$^1$,
S.~Kartal$^2$ \thanks{Corresponding author. E-mail:
sehban.kartal@cern.ch}
, C.~Pagliarone$^3$ \thanks{Corresponding author. E-mail: pagliarone@fnal.gov} \bigskip \\
{\it  $^1$~Benem\'erita Universidad Aut\'onoma de Puebla (BUAP), Puebla, Mexico} \\
{\it $^2$~Istanbul University, Istanbul, Turkey}\\
{\it $^3$~Universit\'a di Cassino \& INFN Pisa, Italy}}
\maketitle

{\large

\begin{abstract}
ACORDE, the ALICE COsmic Ray DEtector is one of the ALICE detectors,
presently under construction. It consists of an array of plastic
scintillator counters placed on the three upper faces of the ALICE
magnet. This array will act as Level 0 cosmic ray trigger and,
together with other ALICE sub-detectors, will  provide precise
information on cosmic rays with primary energies around $10^{15-17}$
eV. In this paper we will describe the ACORDE detector, trigger
design and electronics.
\end{abstract}

\vspace{1.pc}
\section{Introduction} \label{s1}

Because of their extremely low flux, the observation of cosmic ray particles,
with energies around and beyond the knee region, of the cosmic ray spectrum
(see Fig.~\ref{knee}), is possible only using indirect methods.
Ground based arrays, so-called extensive air shower arrays, located on the
Earth surface or underground, detect particle showers created by the interaction
of primary cosmic rays with the atmosphere.
The large collecting surface of such arrays may allow the detection of events
with extremely high energies and very high multiplicity.
One of the main problem of an extensive air shower array experiment,
comes from the difficulty in reconstructing the energy and the mass of the primary
cosmic particle crossing the atmosphere.
As matter of fact a detailed understanding of the interaction mechanism
and and shower development and evolution in the atmosphere is needed in
order to achieve this scientific goal.
On the other hand new phenomena, in very forward high-energy
hadronic interactions, such as coherent pion production,
disoriented chiral condensate states (DCC) or heavy flavor production,
may significantly affect the hadronic cascade and then the
structure of the cosmic ray events at ground level.
Such an effects may be responsible, for example, of the
discrepancies found among different experiments, about the shower
particle composition.
\begin{figure}[t!]
\begin{center}
\includegraphics[width=11cm,height=7cm]{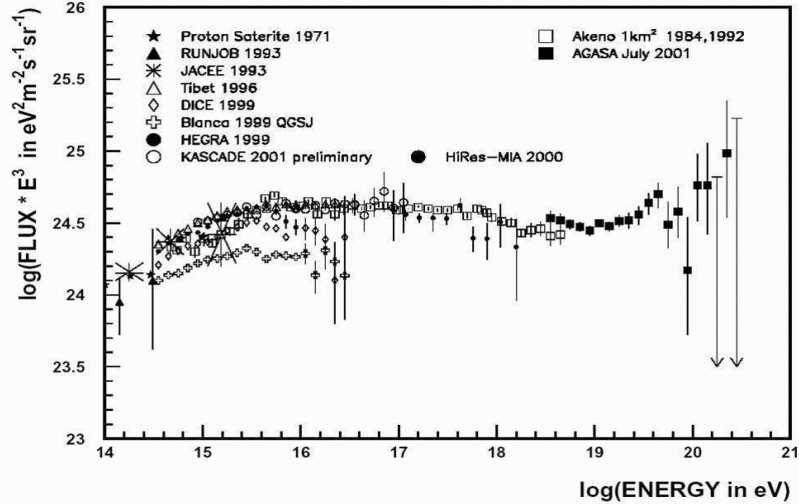}
\vspace{-0.7pc}
\caption{\footnotesize Cosmic ray particle spectrum for charged primaries.
The knee and the ankle are visible near $10^{15}$ and $10^{18.5}$ eV/nucleus~\cite{knee-ref}, respectively.}
\label{knee}
\end{center}
\vspace{-1.7pc}
\end{figure}
Particle production, both at large energies
and in the forward direction, can be estimated only using
models based on the extrapolation of accelerator data.
The interpretation of cosmic ray data, in particular the identification of
primary cosmic rays, rely crucially on such a model.
Presently, there are no accelerator data for particle production
at very small forward regions as well as in the energy region
around the cosmic ray knee.
In the past years, many interaction models such as VENUS, QGSJET, DPMJET, SIBYLL
and others have been extensively developed and used by the cosmic-ray
community~\cite{add1}.
As these models are based on proton-antiproton, electron-positron and
heavy-ion collisions data, at the present, the maximum energy available from
the accelerator experiments comes from the Tevatron collider that has
$ \sqrt{s}=$ $1.96$ TeV.
LHC experiments may then contribute to clarify and constrain such a models
significantly bringing the center-of-mass energy up to $\sqrt{s}=14$ TeV which
corresponds approximately to a proton of $E \simeq 10^{17}$ eV interacting with
an other proton in the upper atmosphere.
This energy value lies above the knee  of the energy spectrum
of cosmic rays, see Fig.~\ref{knee}. ALICE and the other LHC detectors will
allow to extend the study of the $A$-dependence in $AA$ interactions to
higher energies. Particularly important will be in particular to study nitrogen-nitrogen interactions,
since nitrogen and oxygen are the most abundant nuclei in the atmosphere and since
nitrogen nuclei are also present in high-energy cosmic rays ($E \geq 10^{15}$eV).

\begin{figure}[t!]
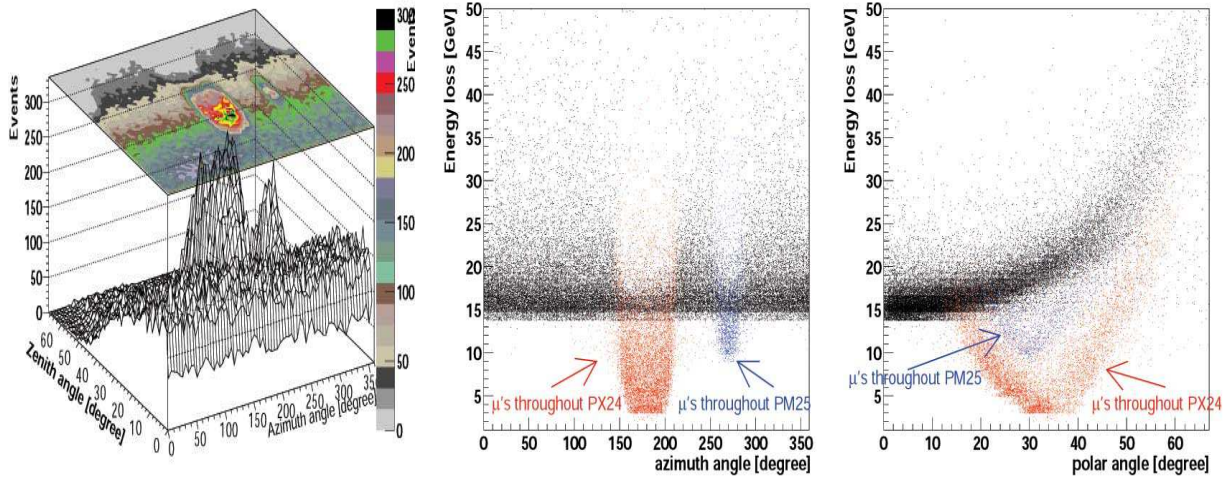

\begin{center}
\includegraphics[width=5.4cm,height=7.cm]{ang1.eps}\hspace{0.6pc}
\includegraphics[width=5.4cm,height=7.cm]{elossvsphi.eps}\hspace{-0.5pc}
\includegraphics[width=5.4cm,height=7.cm]{elossvstheta.eps}
\caption{\footnotesize
a) Angular distribution (zenithal vs azimuthal angle) of atmospheric $\mu$ reaching the ALICE
hall. b) $\phi$ vs Energy loss for atmospheric $\mu$ reaching the ALICE cavern. c)
$\mu$-Energy loss vs (initial) zenith angle.}
\label{phys}
\end{center}
\vspace{-1.pc}
\end{figure}

\noindent
At CERN, the use of large underground high-energy physics experiments, for comic ray studies,
has been suggested by several
groups~\cite{add2}.
In particular, the L3 experiment established a cosmic ray experimental program the so called L3+Cosmics.
The principal aim was to measure, precisely, the inclusive cosmic ray
($\mu$) spectrum in the energy range $20-2000$
GeV~\cite{DETcrt:l3c},
which is relevant for neutrino oscillations.

\section{The ACORDE Detector}

ALICE is an experiment mainly designed to study the products of
nucleus-nucleus collisions at CERN Large Hadron Collider
(LHC)~\cite{ppr}.
The underground location of the ALICE detector, with 30\,m of
overburden composed of subalpine molasse, is an ideal place for muon
based underground experiments (see Fig.~\ref{phys}). Using this facilities, we plan to
observe muon bundles generated by cosmic ray primary particles with
energies around the knee region
$10^{15} \div 10^{17}$ eV~\cite{vienna}.

\noindent
ACORDE, ALICE COsmic Ray DEtector, is an array of scintillator
modules that will act as cosmic ray trigger for ALICE calibration,
as a multiple muon trigger and it will be used to detect, in combination
with other tracking apparatus, atmospheric muons and multi-muons events.
Taking into account the fine granularity of the ALICE Timing Projection
Chamber (TPC) and the fast response of the ACORDE array it will be possible
to measure several properties of cosmic ray events with high density muon
tracks, the so-called muon bundles.
A more detailed discussion on the ACORDE physics goals can be found in the ALICE Physics
Performance Review, Chapter
6~\cite{ppr}.

\subsection{The scintillator array}

ACORDE is an array of plastic scintillator modules (60 at the present)
placed on the top sides of the central ALICE magnet, as shown in
Fig.~\ref{DETcrt_new-array}. More modules, to achieve a better angular coverage and
acceptance, will be added later.
Each module, consists of two plastic scintillator paddles  with
$188\times 20$~cm$^2$ effective area, arranged in a doublet configuration
(see Fig.~\ref{modules}).
Each doublet consists of two superimposed scintillator counters, with their
corresponding photomultipliers active faces, looking back to back.
A coincidence signal, in a time window of $40$ ns, from the two scintillator
paddles gives, for each module, the trigger hit.
A PCI BUS electronics card have been developed in order to measure plateau
and efficiency of the module counters~\cite{scint}.

\begin{figure}[t!]
\begin{center}
\includegraphics[width=8cm,height=8cm]{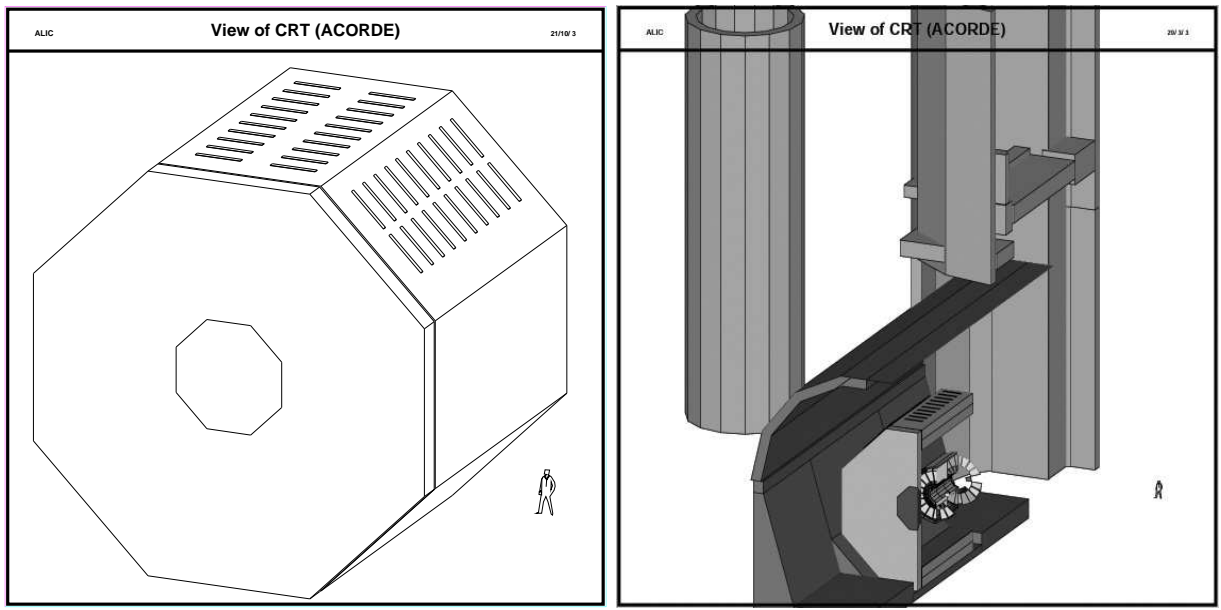}
\includegraphics[width=8cm,height=8cm]{view}
\vspace{-0.4pc}
\caption{\footnotesize a) The ALICE Cosmic Ray Detector; the scintillator array is on top of the ALICE magnet.
b) View of the ALICE cavern.}
\label{DETcrt_new-array}
\end{center}
\vspace{-1.7pc}
\end{figure}

\subsection{The Cosmic Ray Trigger}

The cosmic ray trigger system (CRT) will provide a fast level-zero
trigger signal to the central trigger processor, when atmospheric
muons impinge upon the ALICE detector.
The signal will be useful for calibration, alignment and performance studies of other
ALICE tracking detectors: the Time Projection Chamber (TPC), the Transition Radiation
Detector (TRD)  and the Inner Tracking System (ITS).
A typical rate, for single atmospheric muons crossing the ALICE cavern,
will be less than $1$ Hz/m$^2$.
The expected rate for multi-muon events will be around or less than $0.04$ Hz/m$^2$.
Atmospheric muons will need an energy of at least $17$ GeV to reach
the ALICE hall, while the upper energy limit for reconstructed muons
in the TPC will be less than 2 TeV, depending on the ALICE magnetic
field intensity (up to $0.5$ T).

\begin{figure}[t!]
\begin{center}
\includegraphics[width=8cm,height=6.5cm]{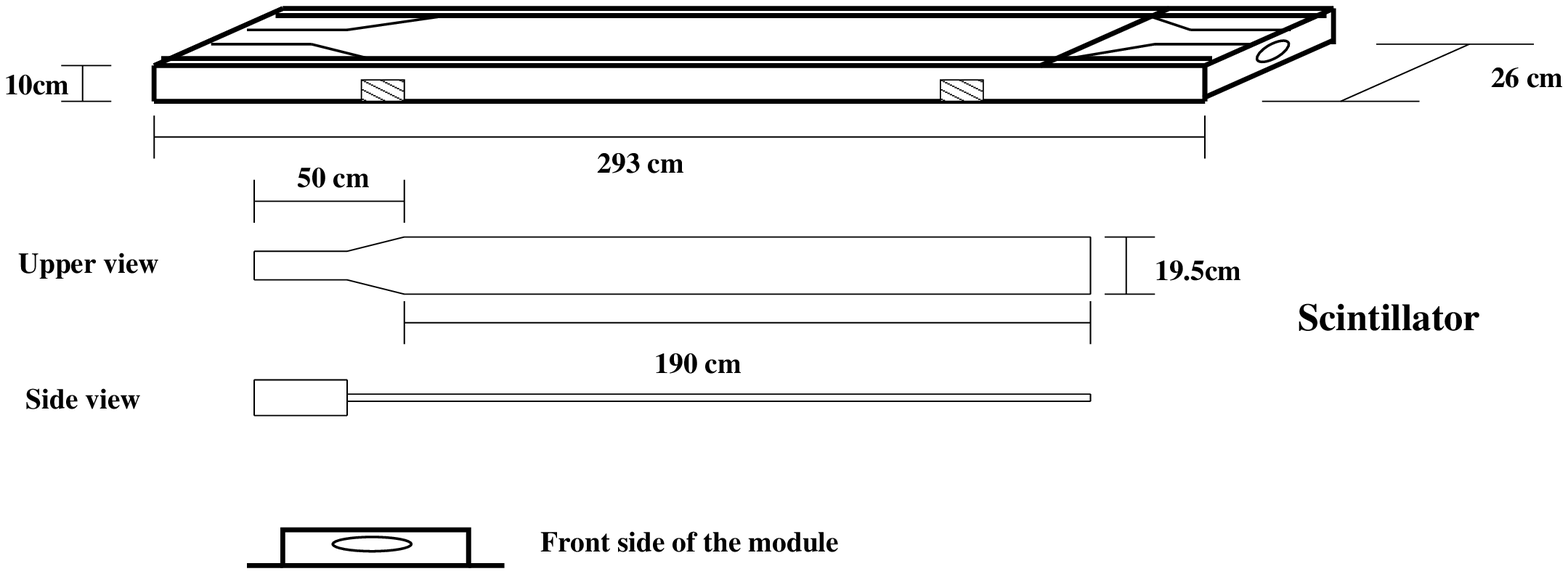}
\includegraphics[width=8cm,height=6.5cm]{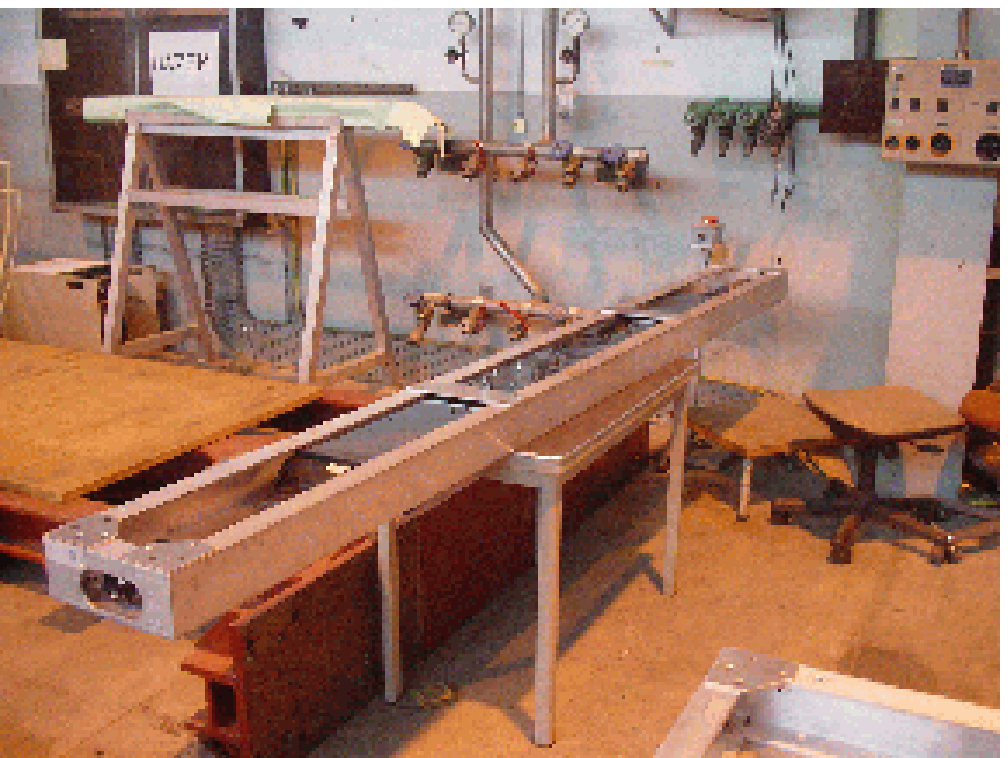}
\vspace{-0.4pc}
\caption{\footnotesize a) Schematic view of an ACORDE module; the
scintillator modules consist of two superimposed scintillator
counters. b) An ACORDE module in the the BUAP test facility.} \label{modules}
\end{center}
\vspace{-1.7pc}
\end{figure}

\subsection{Electronics design and construction}

We have designed and implemented the necessary electronics in order to do the following
tasks:
\begin{enumerate}
\item  Perform the clock synchronization with the LHC clock;
\item  Produce single and multi-muon trigger signals for the Central Trigger Processor (CTP);
\item  Prodce a wake-up signal for the ALICE-TRD;
\item  Communicate with the DAQ through a DAQ Source Interface Unit;
\item  Online monitoring of the scintillator counter performances.
\end{enumerate}

\noindent
The ACORDE Electronics is then made of several different parts essentially the follow:
$i)$ 60 Front End Electronics cards (FEE), one for each ACORDE module;
$ii)$ the ACORDE OR  card to generate the TRD wake up signal (this card
receive the 60 coincidence LVDs signals coming from the FEE cards);
$iii)$ a Main Card, which contains the electronics to receive the 120 LVDS signals coming from 120
scintillators (this card will  produce the single and the multi-coincidence trigger signals and will
provide the connectivity to the ALICE trigger and DAQ systems, see Fig.~\ref{electronics}.a).

\begin{figure}[t!]
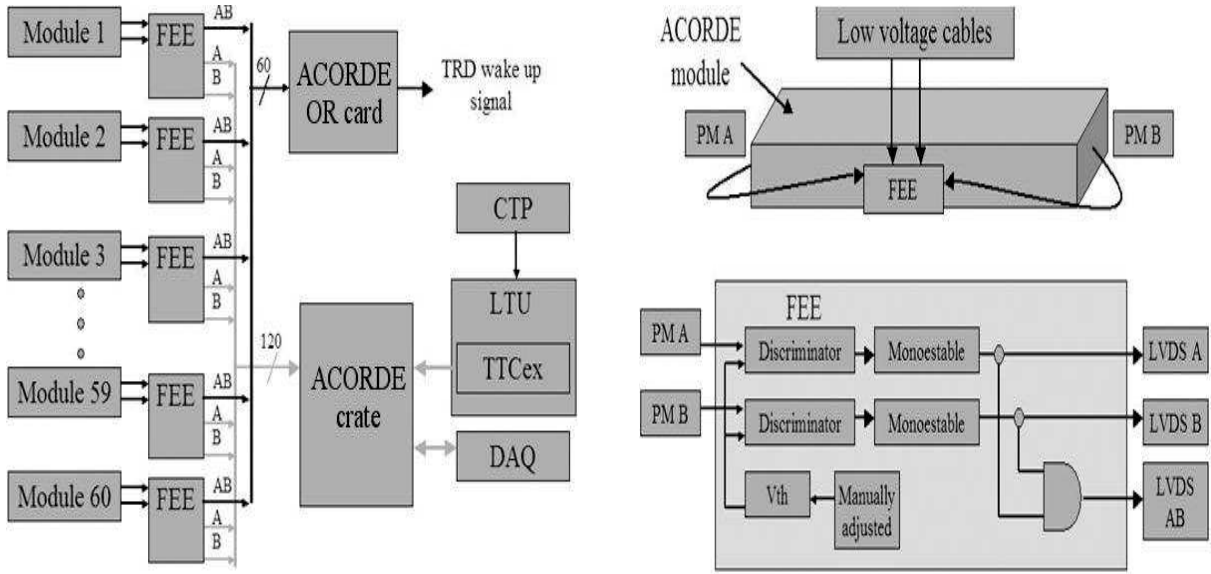

\begin{center}
\includegraphics[width=7.6cm,height=7.6cm]{electro-block.eps}\hspace{1.6pc}
\includegraphics[width=7.6cm,height=7.6cm]{fee-block.eps}
\vspace{-0.4pc}
\caption{\footnotesize a) General scheme of the ACORDE electronics.
b) FEE card block diagram.}
\label{electronics}
\end{center}
\vspace{-1.7pc}
\end{figure}

\noindent
Between the electronics cards the FEE performs the following task:
take the signal from each of the two PMT contained in an ACORDE
module and submits it to a leading edge discriminator. 
The discriminators convert the negative low signal coming from each
PMT to a CMOS digital signal. These signals  have a time duration of
about 10-20ns. In order to produce
a wider signal (100ns) the FEE electronics includes a mono-stable
circuit.
The mono-stable outputs are used to produce the coincidence
(AND) between the two plastic scintillators. Also, the FEE
electronics translate these three signals into LVDS signals to
eliminate noise problems during the transmission to the other cards.
Summarizing, the FEE will be able to provide the coincidence signal
of one ACORDE module and two signals coming from the scintillators,
these three signals will be pulses of 100ns. The coincidence signals
are then transmitted to the ACORDE "OR" card and the PMT signals are
tranfered to the ACORDE main card (see Figure~\ref{electronics}.b).
The ACORDE OR card is used to produce a TRD wake-up signal using the
coincidence signals coming from the 60 FEE cards. Those signals are
then translated from LVDS to CMOS levels. Afterwards, it will be
used an OR gate of 60 inputs to generate the wake-up signal. The
trigger signal is then translated from CMOS to LVDS in order to be
sent to the TRD detector. There is also a 40MHz cryptal oscillator
to produce the wake-up signals (25ns LVDS pulse). The ACORDE OR card
will be located on top of the magnet as shown in
Fig.~\ref{loc-shoe}.a.
\begin{figure}[t!]
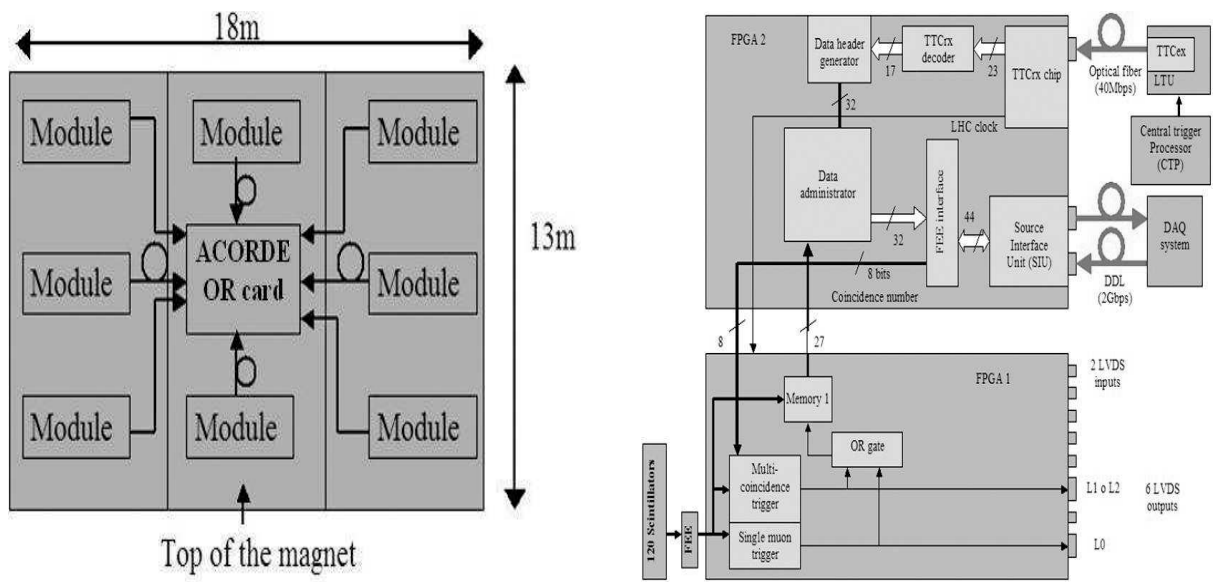

\begin{center}
\includegraphics[width=7.6cm,height=7.6cm]{loc-shoe.eps}\hspace{1.6pc}
\includegraphics[width=7.6cm,height=7.6cm]{electro-mean.eps}
\vspace{-0.4pc}
\caption{\footnotesize a) Location of the ACORDE OR card, on top of the
ALICE magnet. b) Block diagram of the ACORDE main electronics card.}
\label{loc-shoe}
\end{center}
\vspace{-1.7pc}
\end{figure}
The ACORDE main card uses 1 Bus card to interconnect 6 LVDS to CMOS
translator cards with the ACORDE main card. One LVDS to CMOS
translator card is able to receive 20 differential signals coming
from 20 FEE cards. In order to receive the 120 differential signals
we use six translator cards. The Bus card supply the power to the 6
translator cards and the ACORDE main card. The last card receives
120 CMOS signals through the bus card. These signals are supplied to
an ALTERA EP20K100QC240-1 FPGA. This FPGA chip same as the one used
in the TPC electronics; tests done previously have shown that this
device is radiation tolerant. The function of the FPGA chip is to
store the hit information and produce the single muon and the
multi-coincidence trigger in synchronization  with the LHC clock
signal. The main card has a second ALTERA EP20K100QC240-1 FPGA chip,
this device is connected with the TTCrq and SIU card to provide the
LHC clock recovery, the trigger message decoder, the data header
generator and the ACORDE-SIU interface. 
The single moun L0 trigger signal is generated by combinational
logic (AND and OR gates). The ACORDE card latches the tracking data
in a memory when the detector produces a L0 trigger signal. The SIU
daughter card allows the communication with the DAQ system to be
able to send the data header and the tracking information through
the Detector Data Link (DDL). The ACORDE card sends the Data header
and the tracking data to the DAQ system after receiving a L2
accepted trigger message, see Fig.~\ref{loc-shoe}.b. The FEE cards
has been tested showing a very good performance, the leading edge
discriminator has a rise and fall time of 2ns and the mono-stable
circuit produces a 100ns pulse.

\noindent
In order to calibrate the ALICE-TPC detector using cosmic rays, we
have developed a different version of the ACORDE electronics. This
version only receives 40 differential signals coming from 20
modules. Ten modules will be located on top and the other ten will
located underneath. The system will use an ACORDE card with one bus
card to connect two LVDS to CMOS translator cards to 1 main card.
This card will only use one ALTERA EP20K100QC240-1 FPGA chip. We
will not have connection to the Trigger and DAQ ALICE systems. This
electronics will provide a trigger signal (LVDS). The user will be
able to set an arbitrary OR combination of the top modules in AND
with an arbitrary OR combination of the bottom modules through a PC.
The communication between the PC and the ACORDE electronics card is
provided through the  parallel port of the PC. An ACORDE-PC
interface card with a micro-controller has been developed to receive
the data from the PC to show the setting combination with a group of
20 LED indicator. In this way the user can verify if the system have
the correct configuration.

\section{The ACORDE Control System}
The main tasks of any Detector Control System (DCS) is the
monitoring and control of the electrical and physical parameters
necessary to operate the sub-detector. 
In general, these task include the high voltage (HV) and low voltage (LV) power supplies,
the front-end electronics (FEE) configuration parameters, alarm handling,
parameter archiving and all other services such as gas, cooling, etc.
In our case we need to monitor and control only the HV and LV power supplies.
The electronics parameters such as multi-coincidence number, trigger rates
and others will be managed by the central DAQ system.
A general view of our DCS is shown in Fig.~\ref{DCS}.a. We plan
to use the SY1527 and the EASY~\cite{EASY} (Embedded
Assembly SYstem) power supply systems from CAEN as
the HV and LV supplies respectively. The latter is a radiation- and
hig-magnetic-field tolerant power supply system to be installed in the
cavern racks to supply the LV for ACORDE.
The SY1527 crate can hold up to 16 boards. Ten of its slots will be
filled with 12-channel HV boards (A1733N) to feed the 120 PMT's. We
will also install the EASY Branch Controller (A1676A) board that can
control up to 6 EASY crates. We will need two full EASY 3000 crates
each with five 12-channel 2--8 V boards (A3009) for the discriminators.
The remaining four channels for the Acorde Electronics Card and the
shoebox producing the TRD wake up signal will be hosted by a third
EASY 3000 crate with an additional A3009 board.
The intrinsic grouping of the scintillators into modules (each with
a couple of scintillators and PMT's) and the fact that the FEE cards
will be mounted on the modules, makes it natural to select a
detector-oriented hierarchy for our DCS as shown in
Fig.~\ref{DCS}.b. This hierarchy will allow us to independently
control a single module in a simple way which can be useful for
purposes of isolating, testing or calibrating it. 
There will be a total of 183 Control Units (CU) in the DCS; the
parent \verb|ACORDE DCS| with 61 module CU's (M0 through M60). M1 to
M60 will control each one of the modules respectively, including its
associated FEE card. Module~0 will take care of the LV supply to the
Electronics and TRD shoe-box cards.
\begin{figure}[t!]
\begin{center}
\includegraphics[width=7.4cm,height=7.4cm]{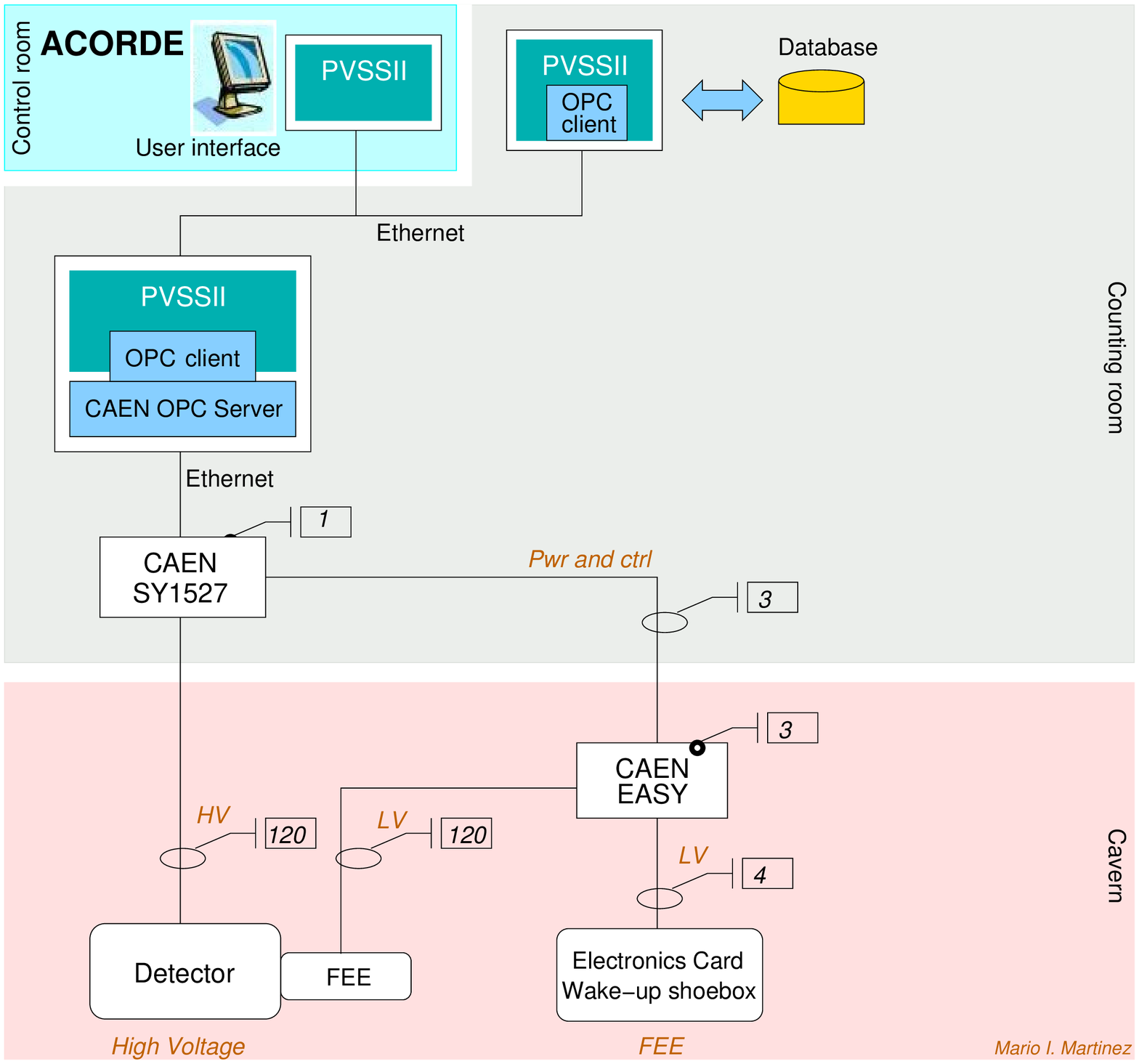}\hspace{1.6pc}
\includegraphics[width=7.4cm,height=7.4cm]{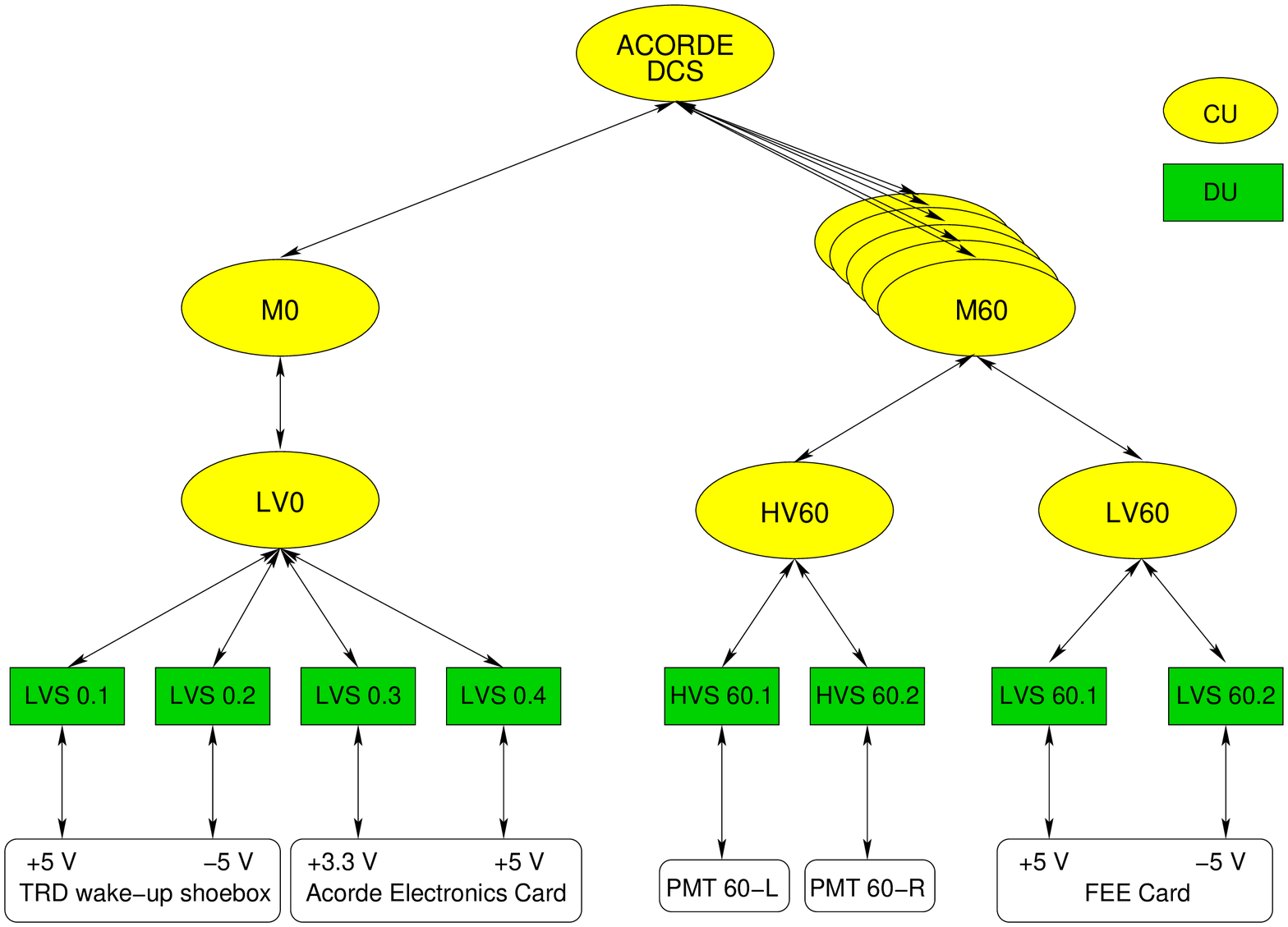}
\vspace{-0.4pc}
\caption{\footnotesize a) Overview drawing of the ACORDE detector control system.
b) Finite state machine hierarchy of the HV and LV of ACORDE.}
\label{DCS}
\end{center}
\vspace{-1.7pc}
\end{figure}
As mentioned before the only electronics parameter to be controlled is
the multi-coincidence number; \emph{i.\ e.}\ the minimum number of
modules hit in order to produce a multi-muon event trigger. This
threshold will be programmed through the DAQ system. The DCS however
should be able to monitor and control it (via the DAQ system). As described
on the ALICE TDR 010~\cite{TDR010} the online systems interface to each
other through the Experiment Control System (ECS); so the DCS will access
the multi-coincidence number by communicating with the ECS.
The DCS is also in charge of managing the calibration procedure of
our detector. This procedure will consist in monitoring the
individual counts on each one of the scintillators in a given time
interval; so that we can realize if any of the corresponding
efficiencies has dropped to an unacceptable value. 

\vspace{-0.4pc}
\section{Summary}
We have implemented a dedicated cosmic ray trigger for ALICE  which
in conjunction with other ALICE detectors provides a powerful tool
for the study of muon bundles properties.
At the present, we have 20 ACORDE modules already installed and the
related electronics working. The ALICE-TPC above ground
commissioning is proceeding  based on  ACORDE trigger using 10
modules placed on the top and 10 underneath the ALICE TPC
(see~\cite{carmine}).

\vspace{-0.4pc}
\section{Acknowledgments}
We are grateful to the  High Energy Physics Latinamerican-European
NETWORK (HELEN) and CONACyT(M\'exico) for continuous support. We
especially thank Laszlo Jenkovszky for his kind hospitality.



}

\end{document}